\pdfoutput=1
\documentclass[11pt]{article}

\usepackage[margin=1in]{geometry}
\usepackage[T1]{fontenc}
\usepackage[utf8]{inputenc}
\usepackage{lmodern}
\usepackage{microtype}
\usepackage{booktabs}
\usepackage{tabularx}
\usepackage{array}
\usepackage{float}
\usepackage{enumitem}
\usepackage{xcolor}
\usepackage{pdflscape}
\usepackage{natbib}
\usepackage{hyperref}
\usepackage{cleveref}
\usepackage{fancyhdr}
\usepackage{setspace}

\definecolor{accent}{HTML}{1F4E79}
\definecolor{lightaccent}{HTML}{EAF1F7}
\definecolor{darkgray}{HTML}{333333}

\hypersetup{
  colorlinks=true,
  linkcolor=accent,
  citecolor=accent,
  urlcolor=accent,
  pdftitle={The Vulnerability With No CVE: Managing Persistent Gaps Between Mandate and Authority in AI Coding Agents},
  pdfauthor={Shayell Aharon Salomon, Amir Shaked, Matan Noga},
  pdfsubject={Position paper proposing a vulnerability-management abstraction for persistent agent-control exposures},
  pdfkeywords={AI coding agents, agentic posture vulnerability, vulnerability management, excessive agency, authorization, least privilege, security posture, runtime governance}
}

\setlist[itemize]{leftmargin=1.35em,itemsep=0.18em,topsep=0.22em}
\setlist[enumerate]{leftmargin=1.45em,itemsep=0.2em,topsep=0.25em}

\newcommand{\term}[1]{\textbf{#1}}

\title{\vspace{-1.3em}\textbf{The Vulnerability With No CVE}\\[0.35em]
\large Managing Persistent Gaps Between Mandate and Authority in AI Coding Agents\\[0.65em]
\normalsize \textit{Position Paper}}
\author{Shayell Aharon Salomon \quad Amir Shaked \quad Matan Noga\\
\small Bluebear Security}
\date{August 2026}

\begin{document}
\maketitle
\vspace{-1.1em}

\begin{abstract}
Existing guidance identifies excessive agency, excessive permission, weak task-bound authorization, and inadequate agent controls as important risks. Control frameworks also describe capabilities for constraining, authorizing, observing, validating, and responding to agent activity. Yet security programs still need a way to manage persistent deployed instances that span components and outlive any one event.

We propose the \emph{agentic posture vulnerability} (APV) as a task-conditioned vulnerability-management abstraction: a durable record for a composed agent-control exposure. One posture may produce different runtime manifestations across tasks; APV links those manifestations to the invariant posture and remains open until authority is narrowed, a missing control is added, risk is accepted, or closure is verified. APV is not proposed as a new root-cause class of risk; it operationalizes existing excessive-agency, authorization, and control-composition weaknesses.

We distinguish APVs from CVE-addressable product defects, OWASP Excessive Agency, Agent Baseline control outcomes, and the runtime authorization--execution gap. We then provide a field vignette, a thresholded definition, six recurring APV patterns, a vulnerability lifecycle, a minimum record, a control-and-closure matrix, tooling implications, and a testable research agenda.
\end{abstract}


\section{Introduction}

Vulnerability management works best when a weakness can be named, located, and remediated. A package has a vulnerable version. A product defect receives an identifier. A scanner maps the flaw to affected assets, a supplier ships a fix, and the security team verifies closure. The CVE program is valuable because it gives tools and teams a shared reference for publicly disclosed product vulnerabilities \citep{cveglossary}.

Not every material weakness is a product defect. Excessive privilege, unsafe defaults, exposed identities, architectural gaps, and security-relevant configuration weaknesses can create exposure even when no software function is broken \citep{nistsp800128}.

For coding agents, the remaining operational question is how to represent a persistent deployed exposure that spans components and outlives any one alert or runtime trajectory. The relevant posture can include the agent harness, model, tools, plugins, connectors, inherited developer identity, credentials, network reach, approval settings, environment boundaries, and telemetry. The weakness may arise from the relationship between what the agent was authorized to accomplish and what it can cause in practice.

A coding agent does not merely hold permission. It interprets a task, inspects state, chooses tools, reacts to failures, and changes course. The same standing posture can therefore appear as a database mutation in one session, a credential transfer in another, and an infrastructure change in a third. The posture stays fixed; the manifestation changes. Individual actions remain detectable, but a rule written for one command may miss the next expression of the same weakness. Detection operates on manifestations; remediation should target the invariant: authority, reach, mediation, and required evidence.

We call that managed security object an \term{agentic posture vulnerability} (APV):

\begin{quote}
A persistent, task-conditioned posture in which a consequential effect is reachable and either exceeds the mandate, lacks required pre-effect mediation, or cannot be attributed and reconstructed to the level needed to govern that authority.
\end{quote}

The \emph{mandate} is not a guess about hidden intent, and it need not exist as one complete pre-written policy. It is an evidence-backed record assembled from explicit user instructions, tickets, repository and organizational policy, environmental constraints, and approved exceptions. For a proposed effect, the available evidence may \emph{support} it, \emph{contradict} it, or leave its status \emph{unknown}. Unknown requires review and is not automatically treated as a violation. Capability approval defines eligible means; it does not authorize every effect for every task.

\emph{Effective authority} is what the agent can cause after tools, credentials, connectors, environment reach, and enforced controls are combined. It is broader than an IAM permission and narrower than every operation a tool merely advertises.

The title is deliberately rhetorical. Agent products can contain ordinary software defects. CVE-2025-55284 affected the confirmation boundary in Claude Code, while CVE-2025-55012 affected permission checks in the Zed Agent Panel; both had affected versions and product patches \citep{anthropic2025cve55284,zed2025cve55012}. APVs address the different case in which the deployed composition creates persistent exposure even when each component behaves as documented.

This paper makes four contributions:

\begin{enumerate}
  \item It separates detections, incidents, runtime authorization divergence, control frameworks, and persistent agentic vulnerabilities as different security objects.
  \item It gives a thresholded operational definition of APV based on persistence, expected tasks, consequential reach, mandate, mediation, and materially necessary evidence.
  \item It identifies six recurring APV patterns and organizes their management in a matrix that overlays established agent-control frameworks.
  \item It proposes a vulnerability lifecycle, a minimum record, and a testable research agenda.
\end{enumerate}

Together, these contributions define a practical unit of work: one durable record for the persistent posture behind changing runtime manifestations.

\section{Related Work and Conceptual Boundary}

The relationship to existing work can be stated simply: OWASP names the risk; Agent Baseline specifies control outcomes; the authorization--execution gap describes runtime divergence; and APV defines the persistent vulnerability record that a security team opens, owns, remediates or accepts, links to runtime evidence, and closes.

Security practice already manages persistent weaknesses through patch and configuration management, while attack-graph research shows how separate conditions can compose into reachable adverse outcomes \citep{nistsp80040r4,nistsp800128,sheyner2002attackgraphs}. OWASP LLM06:2025 Excessive Agency identifies excessive functionality, permissions, and autonomy as root causes, with mitigations that include least privilege, confirmation for high-impact actions, and complete mediation \citep{owasp2025excessiveagency}. APV represents a persistent deployed instance of such risk inside vulnerability management.

Agent Baseline v1.0-draft defines six outcomes: Discover, Constrain, Authorize, Observe, Validate, and Respond \citep{agentbaseline2026}. Its controls cover composition and effective-access mapping, task-bound authority, just-in-time credentials, independent approval, correlated evidence, outcome validation, and response. These controls can prevent, remediate, or provide closure evidence for an APV. The crosswalk used later in this paper is illustrative and version-specific; the APV lifecycle does not depend on Agent Baseline retaining its current structure.

Academic work increasingly treats agent security as an authorization and enforcement problem. Authenticated delegation seeks auditable chains of authority \citep{south2025delegation}; PAuth and intent-governed access control bind authority to concrete task operations \citep{sharma2026pauth,zhu2026igac}; and AuthBench evaluates whether coding agents can infer least-privilege policies \citep{yan2026leastpriv}. AgentSpec, pre-action authorization, AgentBound, and path-based governance enforce policy before or during tool execution \citep{wang2025agentspec,uchibeke2026before,buehler2026agentbound,kaptein2026governance}. Empirical studies also report insecure or benign-but-out-of-scope actions during ordinary coding work \citep{qu2026overeager,kozak2025securitydebt}.

The closest academic concept is the authorization--execution gap (AEG): divergence between what a principal intended to authorize and what an agent executed \citep{wu2026aeg}. AEG describes a runtime trajectory. APV describes a persistent precondition that makes consequential divergence reachable or insufficiently controlled before a particular trajectory occurs. Runtime authorization can remediate an APV, while the vulnerability record preserves scope, ownership, exceptions, linked evidence, and closure. NIST work on agent identity and authorization similarly emphasizes identification, authorization, auditing, and controls; APV focuses on the vulnerability-management layer around those mechanisms \citep{nistagentauth}.

The APV boundary requires a persistent composition, a reasonably expected task class, consequential reach, and a material mandate, mediation, or evidence deficit. This makes the object task-conditioned and applicable to ordinary authorized use, not only adversarial exploitation. It can also exist before any divergent trajectory occurs. Variable manifestations are not unique to agents, but adaptive, task-conditioned execution makes posture-level deduplication and closure especially important. \Cref{tab:objects} summarizes the resulting distinction among the four security objects.

\begin{table}[ht]
\centering
\caption{Related security objects require different responses.}
\label{tab:objects}
\small
\begin{tabularx}{\textwidth}{>{\raggedright\arraybackslash}p{0.19\textwidth} >{\raggedright\arraybackslash}X >{\raggedright\arraybackslash}p{0.20\textwidth}}
\toprule
\textbf{Object} & \textbf{What it represents} & \textbf{Primary response} \\
\midrule
Detection & An observation or signal at a point in time. & Validate and triage. \\
Incident & A bounded event or event chain involving unauthorized action, harm, or an active condition requiring containment or formal response. & Contain and respond. \\
Authorization--execution gap & Runtime divergence between authorization and execution. & Attribute and mediate the divergence. \\
Agentic posture vulnerability & A durable, task-conditioned record of a persistent composed exposure. & Reduce or accept exposure, link evidence, and verify closure. \\
\bottomrule
\end{tabularx}
\end{table}

\section{Motivating Field Vignette}

This position was motivated in part by operational reviews of coding-agent telemetry and configuration. The following vignette is anonymized and is not presented as a prevalence estimate or controlled study.

Three developers on the same team ran coding agents with per-command approvals disabled. Routine confirmations interrupted work, and removing them made the agents faster. The same development environments also contained credentials that could reach production database infrastructure.

During legitimate debugging work, two agents issued \texttt{DROP TABLE} commands against production systems. In both cases the target was a temporary scratch table that the developer had created shortly before and intended to rebuild. Review found the actions authorized and benign; no production data was lost.

The commands were timestamped events and may have produced detections because the operation looked destructive. They were not treated as security incidents under the organization's response criteria because they were authorized and caused no harm. They nevertheless revealed an APV: autonomous execution, production reach, destructive authority, and no independent gate at the point of consequence. No single setting established the vulnerability, and nothing in the control structure distinguished a scratch table from a critical one before execution. Another task could have produced a different production effect. The command revealed the exposure; it did not define it.

An event record captures each command, but not the exposure window during which the composed posture remained reachable. A finding that says only ``approvals disabled'' also loses the production reach, task context, and consequence that determine whether the setting matters.

\subsection{Evidence and Method Note}

The vignette was reconstructed from agent-configuration state, session telemetry, database query logs, and post-event review. The review established that approvals were disabled, production database access was available, the two DDL commands executed, the targets were temporary tables created for the debugging work, the actions matched the intended work, and no production data was lost. The underlying records contain confidential operational details and are not released. The vignette supports construct formation and illustrates the lifecycle; it does not establish prevalence, causality, or independent reproducibility.

\section{An Operational Definition}

\begin{quote}
\textbf{Operational definition.} A posture is an APV when all four conditions hold:
\begin{enumerate}[label=\arabic*.,leftmargin=1.4em,itemsep=0.15em,topsep=0.25em]
  \item The enabling composition persists beyond a single event or trajectory.
  \item It applies to at least one approved or routinely expected class of task.
  \item The agent can reach a consequential effect.
  \item That effect exceeds the mandate, lacks required pre-effect mediation, or cannot be attributed and reconstructed to the level materially required to govern the authority.
\end{enumerate}
\end{quote}

The enabling composition is \term{persistent} when its authority and controls remain until the posture changes; it need not manifest in every task. Expected task classes are approved or routinely anticipated uses of the posture, not arbitrary hypothetical prompts.

A \term{gate gap} means that a consequential effect is reachable without the decision point or enforcement required for that task and environment. A \term{visibility gap} qualifies as an APV only when the agent holds consequential authority and the missing evidence materially prevents attribution, policy evaluation, impact scoping, containment, or verified closure. Generic logging deficiencies remain ordinary control deficiencies.

A control deficiency that is not task-conditioned, does not expose a consequential effect, or has no material bearing on governing agent authority is outside the APV boundary. The definition is intentionally qualitative; it establishes a management threshold rather than a universal numerical score.

The definition makes four boundaries explicit:

\begin{enumerate}
  \item Risk is compositional. A disabled approval, long-lived token, or write-capable connector does not determine severity by itself.
  \item Reachability matters. A reliable block removes an effect from effective authority.
  \item Evidence must be material. Missing telemetry is an APV condition only when it impairs governance of consequential authority.
  \item The planner changes the manifestation, not necessarily the posture. One standing weakness can produce different event signatures across tasks.
\end{enumerate}

Detecting one action cannot, by itself, demonstrate remediation.

\section{Six Recurring APV Patterns}

The following patterns synthesize field observations and established guidance, including OWASP Excessive Agency and Agent Baseline \citep{owasp2025excessiveagency,agentbaseline2026}:

\begin{itemize}
  \item \term{Unsafe agency configuration}: autonomy or approval settings are too permissive for the systems and effects within reach.
  \item \term{Unreviewed installed capability}: plugins, skills, hooks, MCP servers, or extensions expand authority outside the approved inventory.
  \item \term{Over-broad connector authority}: a connector exposes write, administrative, bulk, or workflow effects beyond what the task needs.
  \item \term{Unmediated credential access}: the agent can read or reuse credentials that are broader or longer-lived than the task requires.
  \item \term{Unobserved execution reach}: the agent can act through hosts, services, connectors, or agent types that do not produce the attributable evidence materially required to govern consequential authority.
  \item \term{Collapsed environment boundaries}: routine development sessions can reach production or sensitive systems without an explicit boundary crossing.
\end{itemize}

Across all six, the preferred remediation changes effective authority or the control path. A version update may be necessary when a product defect exists, but it is rarely sufficient for the composed posture.

\section{From Alert Queue to Vulnerability Lifecycle}

Alert-only handling creates a recurring triage loop: an action triggers review, the event is closed, and the same posture later produces another manifestation. Detection rules remain useful, but the vulnerability lifecycle owns the persistent source.

A simple lifecycle is:

\begin{enumerate}
  \item \term{Discover and validate}: join configuration, identity, capability, environment, runtime, and task evidence; confirm that a consequential effect is reachable.
  \item \term{Scope and assign}: identify the affected agents, users, repositories, credentials, connectors, and environments; record the exposure window and owner.
  \item \term{Decide}: remediate, reduce, or accept the risk for a defined period with a named risk owner and review date.
  \item \term{Change the posture}: bound authority, add mediation, restore visibility, separate environments, or broker credentials.
  \item \term{Verify closure}: prove that the vulnerable effect is removed, narrowed, or reliably gated across the affected scope.
\end{enumerate}

A good control turns a possible future incident into an impossible action, a narrower action, or an explicit decision before effect. This does not mean adding prompts everywhere. Poorly placed approvals create fatigue and become click-through theater; strong controls belong where authority becomes consequential.

\subsection{Minimum Vulnerability Record}

An event record is not enough for a standing weakness. \Cref{tab:record} defines the minimum record used throughout the lifecycle.

\begin{table}[H]
\centering
\caption{Minimum record for an agentic posture vulnerability.}
\label{tab:record}
\small
\begin{tabularx}{\textwidth}{>{\raggedright\arraybackslash}p{0.23\textwidth} X}
\toprule
\textbf{Field} & \textbf{Required content} \\
\midrule
Identity and class & Stable identifier, concise title, and one or more control-gap classes. \\
Exposure window and scope & First observed, last verified, affected agents, identities, repositories, connectors, credentials, hosts, and environments. \\
Mandate basis & The task and policy basis used to judge authorized effects, including ambiguity or exceptions. \\
Effective authority and gap & Reachable consequential effects and the missing boundary, gate, credential control, capability restriction, or visibility. \\
Evidence & Configuration snapshots, policy state, runtime observations, reachability evidence, and linked detections or incidents. \\
Owner and disposition & Remediation owner, status, priority, accepted-risk owner, compensating controls, target date, and exception expiry. \\
Remediation & The authority or control change required, not only a new alert rule. \\
Closure evidence & Proof that reach, authority, or control changed and the affected scope was re-verified. \\
\bottomrule
\end{tabularx}
\end{table}

\subsection{Applying the Lifecycle to the Vignette}

Applied to the vignette, the lifecycle joins agent settings, credential reach, session and database telemetry; scopes the affected configurations and production path; assigns owners; changes the posture; and re-tests the affected scope. \Cref{tab:vignette-record} shows the resulting record without inventing client-specific dates or owners.

\begin{table}[H]
\centering
\caption{Illustrative APV record derived from the field vignette.}
\label{tab:vignette-record}
\fontsize{8.4}{9.2}\selectfont
\begin{tabularx}{\textwidth}{>{\raggedright\arraybackslash}p{0.23\textwidth} X}
\toprule
\textbf{Field} & \textbf{Illustrative content} \\
\midrule
Identity and class & Ungated production DDL from development agents; unsafe agency configuration and collapsed environment boundaries. \\
Exposure window and scope & Three developer-agent configurations, from the first verified coexistence of disabled approvals and production database reach until either condition changed. Exact dates are omitted. \\
Mandate basis & Authorized debugging work. The record would preserve whether production DDL was explicitly requested; capability availability alone would not settle the question. \\
Effective authority and gap & Production DDL was reachable through inherited credentials, with no independent gate that distinguished a scratch target from a critical one. \\
Evidence & Agent settings, session telemetry, database logs, post-event review, and the two benign \texttt{DROP TABLE} events linked to the posture. \\
Owner and disposition & Named platform, identity, and database-control owners; remediate or accept for a defined period with a review date. \\
Remediation & Read-only production access by default, just-in-time elevation, and step-up authorization before production DDL. \\
Closure evidence & Re-test every affected configuration and confirm that production DDL is denied or independently gated; record the verified scope and date. \\
\bottomrule
\end{tabularx}
\end{table}

\clearpage
\begin{landscape}
\section{The APV Control and Closure Matrix}

The APV Control and Closure Matrix organizes five generic vulnerability-management functions across the six recurring patterns. Its structure follows the design principle of the Cyber Defense Matrix \citep{yu2026cdm}. \Cref{tab:matrix} applies Discover, Bound, Mediate, Observe, and Verify to each pattern.

The functions have a close, version-specific crosswalk to Agent Baseline v1.0-draft \citep{agentbaseline2026}: Discover maps to Discover, Bound to Constrain, Mediate to Authorize, Observe to Observe, and Verify to Validate. The crosswalk is illustrative; the APV lifecycle does not depend on that framework retaining its current structure. Agent Baseline's Respond outcome addresses active unsafe activity, which not every standing vulnerability requires.

\begin{table}[H]
\centering
\caption{APV Control and Closure Matrix: representative objectives by APV pattern and management function.}
\label{tab:matrix}
\small
\setlength{\tabcolsep}{4.4pt}
\renewcommand{\arraystretch}{1.42}
\begin{tabularx}{0.95\linewidth}{>{\raggedright\arraybackslash}p{0.155\linewidth} *{5}{>{\raggedright\arraybackslash}X}}
\toprule
\textbf{APV pattern} & \textbf{Discover} & \textbf{Bound} & \textbf{Mediate} & \textbf{Observe} & \textbf{Verify} \\
\midrule
Unsafe agency configuration & Inventory autonomy by environment. & Restrict high-impact actions. & Require approval before consequential effects. & Monitor configuration and actions. & Re-test that actions are blocked or gated. \\
\addlinespace
Unreviewed installed capability & Inventory capabilities and provenance. & Allowlist and scope capabilities. & Approve installation and sensitive use. & Log loading and invocation. & Confirm revoked capabilities cannot run. \\
\addlinespace
Over-broad connector authority & Map targets and write scope. & Restrict write operations. & Gate administrative and irreversible calls. & Record target, identity, and effect. & Test that out-of-scope calls are denied. \\
\addlinespace
Unmediated credential access & Map credential scope and lifetime. & Use short-lived, task-scoped credentials. & Require just-in-time issuance. & Correlate issuance and use. & Verify revocation and expiry. \\
\addlinespace
Unobserved execution reach & Map reach against telemetry coverage. & Block unmonitored execution paths. & Route sensitive actions through monitored gateways. & Preserve attributable evidence. & Confirm no consequential blind path remains. \\
\addlinespace
Collapsed environment boundaries & Map development-to-production paths. & Separate identities and default to read-only. & Require explicit boundary crossing. & Record environment transitions. & Test that production writes require the gate. \\
\bottomrule
\end{tabularx}
\end{table}
\end{landscape}
\clearpage

The columns are not interchangeable. Better observation may improve investigation while leaving the authority mismatch intact. Because manifestations vary, closure cannot be demonstrated by the disappearance of one command pattern. Closure requires evidence across the affected scope that the consequential effect is no longer reachable, is narrower, is reliably mediated, or is covered by an explicit risk decision.

\section{Implications for Security Tooling}

Conventional tools see important parts of agentic risk: endpoint monitoring sees commands, IAM sees permissions, cloud security sees production access, secret scanners see credentials, database monitoring sees queries, and the agent platform sees approval settings.

The missing layer is a task-conditioned representation that connects mandate, effective authority, enforced controls, reachability, runtime evidence, and closure. The data model should support both directions: heterogeneous detections and incidents may be evidence of one APV, while one APV may anticipate several classes of manifestation. An APV-aware program therefore connects configuration and capability inventory, reviewable task and policy context, runtime evidence, and a posture lifecycle with ownership, exceptions, remediation, and closure. This complements EDR, IAM, CIEM, CSPM, DLP, secret scanning, database monitoring, control frameworks, and runtime authorization.

\section{Testable Research Agenda}

The framework should be evaluated rather than accepted as terminology alone. The following hypotheses are not results; they state measurable tests that could support or falsify the position:

\begin{enumerate}
  \item \term{Compositional severity}. Across labeled posture scenarios, a model using posture composition will produce fewer pairwise ranking inversions against expert-adjudicated risk rankings than a model based on one setting alone.
  \item \term{Stable posture, different manifestations}. Holding posture constant, variation across approved task classes will yield more distinct consequential action classes than repeated runs within one task class.
  \item \term{Root-control leverage}. Across a fixed task suite, an authority-reducing or independently mediating control will prevent more distinct risky action classes than a rule matching one observed manifestation.
  \item \term{Event-model mismatch}. Compared with alert-only handling, an APV workflow will show a higher owner-assignment rate and more verified closures, together with lower recurrence and shorter median exposure duration.
\end{enumerate}

These outcomes can be measured through paired posture scenarios, repeated runs across task families, intervention studies comparing root controls with detection-only controls, and longitudinal studies of security operations. Further work is also needed on mandate construction, indirect tool chains, dynamic credential discovery, and benchmarks that report which posture made an unsafe action reachable.

A practical interim approach to prioritization can consider impact, reachable systems, autonomy, affected scope, observability, persistence, distance between expected tasks and effective authority, and observed manifestations. This paper does not claim a validated universal score.

\section{Limitations}

This is a position paper, not a prevalence study. The field vignette motivates the construct but does not establish how common APVs are, and its confidential operational evidence cannot be independently reproduced. The six patterns overlap and are not exhaustive.

APV builds on Excessive Agency, misconfiguration, excessive entitlement, attack paths, and established control frameworks rather than introducing a new root-cause class of risk. Its value depends on whether the task-conditioned record improves ownership, deduplication, remediation, risk acceptance, linkage to runtime evidence, and verified closure. The current Agent Baseline crosswalk reflects v1.0-draft and should be revisited when that framework stabilizes.

The boundary between an APV and an ordinary control deficiency remains partly judgment-based. We require persistence, a reasonably expected task class, consequential reach, and a material mandate, mediation, or evidence deficit. The threshold needs empirical validation, especially for visibility-only cases.

Mandate evidence can remain incomplete or contested, and task scope can legitimately change. The supported, contradicted, and unknown states require policy precedence and human review; an inferred intent should not automatically become authority. Effective authority can also be difficult to enumerate when agents discover tools and credentials, invoke nested services, delegate to sub-agents, or change the environment during execution.

Product defects and control bypasses should still receive CVEs and patches when an affected product and fix exist. APV addresses the different case in which exposure is composed across deployed systems and persists until authority or control changes.

\section{Conclusion}

AI coding agents turn familiar configuration, entitlement, and control posture into adaptive execution. The operational challenge is to manage a persistent deployed exposure when it spans components, survives the end of a session, and may appear differently in the next task.

APV gives security teams one durable, task-conditioned vulnerability record for that posture. It links runtime evidence, preserves scope and ownership, and remains open until authority is narrowed, a missing control is added, risk is accepted, or closure is verified.

The practical question is therefore not only whether code and dependencies are secure. It is also:

\begin{quote}
\textbf{What effective authority does each agent have for this task, behind which controls, and with what visibility?}
\end{quote}

There is no single version bump for giving an agent too much room. There is the work of making that room explicit, bounded, observable, and enforceable before the agent decides how to use it.

\section*{Data Availability}

The operational records underlying the field vignette cannot be shared because they contain confidential organizational information. The paper provides an anonymized account, an evidence description, and an illustrative vulnerability record.

\section*{Competing Interests}

The authors are affiliated with Bluebear Security, which develops security technology for AI coding agents. This paper presents a general framework and does not evaluate any commercial product.

\begingroup
\small
\setlength{\bibsep}{0.25em}
\bibliographystyle{plainnat}
\bibliography{references}
\endgroup

\end{document}